\documentclass[12pt]{article}

\usepackage{amsfonts}
\usepackage{amsmath}
\usepackage{amssymb}
\usepackage{mathtools}

\usepackage[english]{babel}
\usepackage{hyperref}
\usepackage[indent]{parskip}

\usepackage{amsthm}

\theoremstyle{definition}

\theoremstyle{remark}
\newtheorem*{remark}{Remark}

\theoremstyle{example}
\newtheorem{example}{Example}[section]

\numberwithin{equation}{section}

\renewcommand{\thefootnote}{\fnsymbol{footnote}}

\DeclareMathOperator{\mathd}{d}
\DeclareMathOperator{\mathe}{e}
\DeclareMathOperator{\mathi}{i}
\DeclareMathOperator{\Aut}{Aut}

\DeclareMathOperator{\mcdp}{McdP}
\DeclareMathOperator{\mcdq}{McdQ}
\DeclareMathOperator{\supermacdonald}{SMcdP}
\DeclareMathOperator{\schur}{Schur}
\DeclareMathOperator{\id}{Id}

\DeclareMathOperator{\li}{Li}

\newcommand\FI{\nu}

\newcommand\shift[2]{T_{#1,#2}}
\newcommand\ts{{\bf t}}
\newcommand\oper{\mathcal{D}}
\newcommand\langlands{{{}^{\mathrm{L}}\omega}}
\newcommand\littleQJacobi{\mathcal{P}}
\newcommand\staircase{{\boldsymbol\delta}}

\begin{document}
\pagenumbering{gobble}
\noindent{\hfill\small\texttt{UUITP-32/21}}\par\smallskip\vskip24pt
\begin{center}
 \Large
 {\bf{On refined Chern-Simons and refined ABJ matrix models}}
 \\[12mm] \normalsize
 {\bf Luca Cassia\footnote{\texttt{luca.cassia@physics.uu.se}} and Maxim Zabzine\footnote{\texttt{maxim.zabzine@physics.uu.se}}} \\[8mm]
 {\small\it
  Department of Physics and Astronomy,
  Uppsala University,\\
  Box 516,
  SE-75120 Uppsala,
  Sweden\\
  \vspace{.5cm}
 }
\end{center}
\vspace{7mm}
\begin{abstract}
We consider the matrix model of $U(N)$ refined Chern-Simons theory on $S^3$ for the unknot. We derive a $q$-difference operator whose insertion in the matrix integral reproduces an infinite set of Ward identities which we interpret as $q$-Virasoro constraints. The constraints are rewritten as difference equations for the generating function of Wilson loop expectation values which we solve as a recursion for the correlators of the model.
The solution is repackaged in the form of \textit{superintegrability} formulas for Macdonald polynomials. Additionally, we derive an equivalent $q$-difference operator for a similar refinement of ABJ theory and show that the corresponding $q$-Virasoro constraints are equal to those of refined Chern-Simons for a gauge super-group $U(N|M)$. Our equations and solutions are manifestly symmetric under Langlands duality $q\leftrightarrow t^{-1}$ which correctly reproduces 3d Seiberg duality when $q$ is a specific root of unity.
\end{abstract}
\vfill

\eject
\normalsize

\renewcommand{\thefootnote}{\arabic{footnote}}
\setcounter{footnote}{0}

\newpage
\pagenumbering{arabic}
\tableofcontents

\parskip=4pt

\section{Introduction}

In the seminal paper \cite{Witten:1988hf}, Witten gave a gauge field theory construction of the celebrated Jones polynomials for knots and links in 3 dimensions using the path integral of Chern-Simons (CS) theory. Since then many generalizations of this important result have been worked out, both from the mathematical side, by defining knot (co)homologies that give categorifications of the knot polynomials, and also from the physics side where people have been able to compute exactly some BPS observables in certain supersymmetric versions of CS theory by using localization techniques.

In \cite{Aganagic:2011sg}, Aganagic and Shakirov used a deformation of the Verlinde ring of $\widehat{su}(N)_k$ to give a new physical interpretation for certain refined knot invariants that were previously constructed using homological methods in \cite{Dunfield:2005si}. More specifically, using $M$-theory they conjectured a new type of ``refined'' gauge theory in 3 dimensions whose Wilson-loop observables can be identified with the Poincar\'e polynomials of Khovanov-Rozansky HOMFLY homology \cite{Khovanov:2000cat,Khovanov:2008roz}. Their new theory can be seen as a deformation of CS gauge theory known as \textit{refined Chern-Simons} (rCS) which depends on two parameters $q$ and $t$ that appear in the definition of the modular $S$ and $T$ matrices. In the case of usual CS these matrices are written in terms of characters of the gauge group, which in the case of $U(N)$ are Schur symmetric polynomials. In the refined theory one substitutes Schur polynomials with their $q,t$ deformation known as Macdonald polynomials which are know to play an important role in the theory of symmetric functions and in representation theory as well.

In their paper they also gave a matrix model definition of the partition function of rCS on $S^3$ arguing that (unknot) Wilson-loop observables are given by averages of Macdonald functions inside of the matrix integral. This matrix model falls into the category of $q,t$-deformed matrix models for which it was shown there exists an action of the deformed Virasoro algebra \cite{Shiraishi:1995rp} in the form of Ward identities known as $q$-Virasoro constraints \cite{Nedelin:2015mio}.

In this paper we explicitly derive the $q$-Virasoro constraints for the matrix model of rCS theory on $S^3$ by making use of insertions of certain finite difference operators acting on the integration variables of the matrix integral. The constraints that we derive can be rewritten as a system of infinitely many linear relations between the correlators of the model which in turn can be packaged into some difference equations for the generating function of the matrix model expectation values.

An explicit analysis of these equations reveals that they can be integrated completely to give a full solution of the model in terms of normalized correlation functions, which is equivalent to the statement that the relations between these correlators form a recursion which is determined uniquely by the initial data provided by the empty correlator, which is identified with the partition function itself.
Computing the solution up to some finite order we are able to conjecture a closed formula for the average of Macdonald characters, and we find that it can be nicely written in terms of the same characters evaluated at some specific locus. This shows that rCS matrix model enjoys a property know as \textit{superintegrability} similarly to what was shown to happen for other $q,t$-deformed matrix models \cite{Cassia:2020uxy}.

Similarly to the case of ordinary unitary gauge groups $U(N)$, one can also define CS theory for unitary super-groups $U(N|M)$ and in fact there exists a refinement deformation of this super-group gauge theory as well. This theory coincides with the refinement of the ABJ theory of \cite{Aharony:2008ug,Aharony:2008gk}. The corresponding matrix model is therefore the supersymmetric generalization of the rCS one and we show that it also admits an action of the $q$-Virasoro algebra. Moreover, we show that the constraints can be written in the same way as those of rCS once we substitute the rank $N$ with the \textit{effective} rank $N-\tfrac{\log q}{\log t}M$. By using this observation, it follows that the constraints in the two cases have essentially the same solution up to the choice of overall normalization and rank.

Finally, we show that the refined matrix model representation of the $q$-Virasoro algebra admits a non-trivial involution which exchanges $q\leftrightarrow t^{-1}$ known under the name of \textit{quantum $q$-geometric Langlands} duality, and it acts non-trivially on the correlators in such a way that the full (normalized) generating function is left invariant. The physical interpretation of such a symmetry can be traced back to a 3d version of Seiberg duality which for pure CS reduces to the well-known level-rank duality of the representation ring of $\widehat{u}(N)_k$

\section{Refined Chern-Simons matrix model}
\label{sec:rcsmm}

We start by recalling the definition of the refined CS matrix model as given in \cite{Aganagic:2011sg,Aganagic:2012ne}. The refinement is obtained by introducing a specific $\beta$-deformation of the standard CS matrix model of \cite{Marino:2002fk,Aganagic:2002wv} on the 3-sphere.

\subsection{The partition function}

The theory is described by the following data: the CS level $k\in\mathbb{Z}$, the gauge group which in our case we take to be $U(N)$ so that the rank and the dual Coxeter number are both $N$. Additionally one can consider introducing a Fayet–Iliopoulos (FI) term $\FI$ for the $U(1)$ center of the gauge group.

The unknot partition function of unrefined CS is given by the Stieltjes–Wigert matrix integral
\begin{equation}
\label{eq:unrefined-CS}
\begin{aligned}
 Z_N^\mathrm{CS}(q) &= \frac{(-1)^{N(N-1)/2}}{N!} \int_{\mathbb{R}^N}
 \prod_{1\leq i < j\leq N} \left(2\sinh\frac{z_i-z_j}{2}\right)^2
 \prod_{i=1}^N \mathe^{-\frac{z_i^2}{2\log q}} \mathd z_i \\
 &= \frac{1}{N!} \int_{\mathbb{R}_+^N}
 \prod_{i \neq j} \left(1-x_i/x_j\right)
 \prod_{i=1}^N \mathe^{-\frac{\log^2 x_i}{2\log q}} \frac{\mathd x_i}{x_i}
\end{aligned}
\end{equation}
where in the second line we adopted the change of variables $z_i=\log x_i$. The matrix integral depends explicitly on the rank $N$ of the matrices but also on a complex parameter $q$ which appears in the denominator of the potential function. In order to identify the matrix model partition function with the gauge theory partition function one usually assumes that $q$ is of the form
\begin{equation}
\label{eq:q-level}
 q=\mathe^{\frac{2\pi\mathi}{k+N}}
\end{equation}
however, we observe that the matrix integral, when convergent, can be regarded as a well-defined function over the complex plane and not just at roots of unity. 

The refinement of CS theory corresponds to a specific type of deformation of the matrix integral \eqref{eq:unrefined-CS}, where one substitutes the classical Vandermonde measure with a $q,t$-analogue of it, i.e. the function
\begin{equation}
 \Delta_{q,t}(x) = \prod_{i\neq j} \frac{(x_i/x_j;q)_\infty}{(tx_i/x_j;q)_\infty}
\end{equation}
where $(x;q)_\infty = \prod_{n=0}^\infty (1-q^n x)$ is the $q$-Pochhammer symbol.
The refined partition function can then be written as
\begin{equation}
\label{eq:partition-function-rCS}
\begin{aligned}
 Z_N^{\mathrm{rCS}}(q,t)
 &= \frac{1}{N!}\int_{\mathbb{R}^N_{+}}
 \Delta_{q,t}(x) \prod_{i=1}^N \mathe^{-V(x_i)} \mathrm{d}x_i
\end{aligned}
\end{equation}
where the potential $V(x)$ is given by the function
\begin{equation}
\label{eq:potential-rCS}
 V(x_i) = \frac{\log^2x_i}{2\log q} + (-\beta(N-1)-\FI)\log x_i
 \quad\quad\quad
 \text{with}
 \quad\quad\quad
 \beta = \frac{\log t}{\log q}
\end{equation}
which now also includes an arbitrary FI term proportional to $\log x_i$. When $t=q$ (i.e. $\beta=1$) the function $\Delta_{q,t}(x)$ reduces to the CS measure $\prod_{i\neq j}(1-x_i/x_j)$ and the partition function in that limit can be identified with the integral in \eqref{eq:unrefined-CS} provided we choose the FI parameter as $\nu=-N$.
For later convenience we also introduce the additional parameters
\begin{equation}
\label{eq:parameters}
 r= q^\nu~,
 \quad\quad\quad
 p= qt^{-1}~.
\end{equation}

The matrix integral \eqref{eq:partition-function-rCS} depends now on the parameters $q$ and $t$ (and $r$) which we regard as arbitrary complex numbers. As before, the identification with the refined gauge theory requires that we choose specific values for these parameters compatible with the fact that the $S$ and $T$ matrices provide a unitary representation of $SL(2,\mathbb{Z})$ on the Hilbert space of a 2-torus. This happens when $q$ and $t$ are given by
\begin{equation}
\label{eq:qt-level}
 q=\mathe^{\frac{2\pi\mathi}{k+\beta N}}~,
 \quad\quad\quad
 t=\mathe^{\frac{2\pi\mathi\beta}{k+\beta N}}
\end{equation}
which is the refinement of the relation \eqref{eq:q-level}, however we remark that the matrix model itself does make sense for arbitrary $q$ and $t$.

\begin{remark}
The partition function $Z_N^{\mathrm{rCS}}(q,t)$ is a convergent integral over the positive real line provided $|q| > 1$. This can be seen most easily in the variables $z_i=\log x_i$ in which case we have a convergent real Gaussian integral over the contour $\mathcal{C}=\mathbb{R}$. For $|q|\leq 1$ the integrand does not fall-off at infinity on the real line, however one can make the integral convergent by modifying the integration contour. For instance, for $|q|=1$ one should rotate the contour slightly off the real axis while for $|q|<1$ one can take $\mathcal{C}=\mathi\mathbb{R}$.
Furthermore, one can even substitute the Riemann integral $\int\mathd x_i$ with the discrete Jackson integral $\int\mathd_q x_i$ without affecting the derivation of the $q$-Virasoro constraints as discussed in \cite{Lodin:2018lbz}, however we do not have intuition for a physical theory that would give rise to such a ``discrete'' matrix model.
\end{remark}

The partition function of the refined matrix model computes the vacuum matrix element of the product $TST$
\begin{equation}
 Z_N^{\mathrm{rCS}}(q,t) = \langle 0|TST|0\rangle
\end{equation}
where $S$ and $T$ are Macdonald deformations of the Kac-Peterson modular matrices, first introduced in \cite{Kirillov:1995k,Cherednik:1995}.
This matrix element corresponds to the path integral of rCS theory on $S^3$ with no Wilson loops.

As argued in \cite{Aganagic:2011sg,Aganagic:2012ne}, given a representation $R_\lambda$ of the gauge group $U(N)$ one can compute the insertion of the corresponding unknot Wilson loop operator by inserting the corresponding character in the matrix integral in \eqref{eq:partition-function-rCS}. In the unrefined case, the $U(N)$ characters are given by Schur polynomials in the integration variables $\{x_i\}$, while in the refined case these are substituted by Macdonald symmetric polynomials $\mcdp_\lambda(x;q,t)$ {or $\mcdq_\lambda(x;q,t)$ (see Appendix~\ref{sec:symmetric-functions})}. The corresponding entry of the modular $TST$-matrix then computes the Macdonald expectation value
\begin{equation}
 \langle R_\lambda|TST|0\rangle = \langle\mcdp_\lambda(x;q,t)\rangle
\end{equation}
{and similarly for $\mcdq_\lambda$ using \eqref{eq:mcdPQ}.}
Observe that the Wilson loop wraps around the non-contractible cycle of one of the two solid tori that glue to give the 3-sphere. Moreover, one can interpret the action of the $T$-matrix as shifting the framing of the unknot by 1 unit.

Similarly, one can compute the rCS path integral of an Hopf link by evaluating the expectation value of the product of two Macdonald polynomials, however in this paper we will mostly focus on the unknot partition function.

\subsection{Generating function of correlators}

The correlation functions of the matrix model are the expectation values of products of power-sums $p_n(x)=\sum_{i=1}^N x_i^n$ (see Appendix \ref{sec:symmetric-functions}),
\begin{equation}
 c_\lambda = \left\langle p_\lambda(x) \right\rangle
 = \frac{1}{N!}\int_{\mathbb{R}^N_{+}}
 \Delta_{q,t}(x)
 \sum_{i_1,\dots,i_N=1}^N x_{i_1}^{\lambda_{1}}\dots x_{i_N}^{\lambda_{N}}
 \prod_{i=1}^N \mathe^{-V(x_i)} \mathrm{d}x_i
\end{equation}
where $\lambda=[\lambda_1,\dots,\lambda_N]$ is an integer partition of length at most $N$. It is useful to repackage all the correlators $c_\lambda$ into a generating function. A natural way to do this is to take the expectation value of the generating function of all power-sums
\begin{equation}
\label{eq:reproducing-kernel-rCS}
 \Pi(x,\ts;q,t)
 = \exp\left(\sum_{n=1}^\infty \frac{t_n}{n}\frac{1-t^n}{1-q^n}p_n(x)\right)
 = \sum_{\lambda} z_\lambda(q,t)^{-1} p_\lambda(x) t_\lambda
\end{equation}
with
\begin{equation}
 t_\lambda = \prod_{i=1}^N t_{\lambda_i}
\end{equation}
and $z_\lambda(q,t)$ defined as in \eqref{eq:z-lambda-qt}.
The function $\Pi(x,\ts,q,t)$ is a formal power series in the auxiliary variables $\ts=\{t_1,t_2,\dots\}$ which are usually called \textit{higher times}.
If we regard the higher times $t_n$ as the power-sums of a second set of independent variables, then $\Pi(x,\ts,q,t)$ can be interpreted as the reproducing kernel for the Macdonald inner product $(\cdot,\cdot)_{q,t}$ on the ring of symmetric functions $\Lambda_{q,t}$.

The generating function of all correlators is defined as
\begin{equation}
\label{eq:generating-function-rCS}
 Z_N^{\mathrm{rCS}}(\ts;q,t) = \left\langle\Pi(x,\ts;q,t)\right\rangle
 = \sum_{\lambda} z_\lambda(q,t)^{-1} c_\lambda t_\lambda
\end{equation}
Observe that because $\Pi(x,\ts;q,t)$ is a reproducing kernel, then the following identity holds
\begin{equation}
\label{eq:adjointpowersum}
 p_n(x) \Pi(x,\ts;q,t) = t_n^\perp \Pi(x,\ts;q,t)
\end{equation}
where $t_n^\perp$ is the adjoint of multiplication by $t_n$, defined by
\begin{equation}
\label{eq:adjoint-power-sum}
 t_n^\perp = n\frac{1-q^n}{1-t^n} \frac{\partial}{\partial t_n}~,
 \quad\quad\quad
 n\geq1
\end{equation}
With these definitions we can then write all correlators as
\begin{equation}
 c_\lambda
 = t_\lambda^\perp Z_N^{\mathrm{rCS}}(\ts;q,t)|_{\ts=0}
\end{equation}
Moreover, the operators of multiplication by times $t_n$ and their adjoints $t_n^\perp$ form an Heisenberg algebra with generators
\begin{equation}
\label{eq:heisenberg}
 a_{n} = \left\{
 \begin{array}{ll}
  t^\perp_{n}&n\geq 1 \\
  N&n=0 \\
  t_{-n} & n\leq-1
 \end{array}\right.
\end{equation}
and commutation relations
\begin{equation}
 [a_m,a_n] = m\frac{1-q^{|m|}}{1-t^{|m|}}\delta_{m+n,0}
\end{equation}

\subsection{q-Virasoro constraints}
\label{sec:q-virasoro-constraints}

The $q$-Virasoro constraints for the matrix model in \eqref{eq:partition-function-rCS}
can be obtained by introducing an appropriate finite difference operator in the integral of the generating function. As in the case of the classical Virasoro constraints, the insertion of the difference operator leads to a vanishing integral even though the integrand itself is not identically zero. Rewriting the integral as a linear combination of correlators gives a non-trivial set of relations between these functions.

In order to define the difference operator appropriate for the rCS matrix model we first introduce the shift operator
\begin{equation}
 \shift{q}{x_i} f(x_1,\dots,x_i,\dots,x_N) = f(x_1,\dots,qx_i,\dots,x_N)
\end{equation}
which acts by rescaling the variable $x_i$ by the factor $q$. Next we define the $q$-derivative by
\begin{equation}
 D_{q,x_i} f(x) = \frac{1-\shift{q}{x_i}}{(1-q)x_i}f(x)
\end{equation}
and finally we can define our difference operator as
\begin{equation}
\label{eq:q-diff-operator}
 \oper_{m} f(x) = \sum_{i=1}^N D_{q,x_i}
 \left[q^{-\frac{m}{2}} x_i^{m+1}A_i(x;t^{-1})f(x)\right]
\end{equation}
where
\begin{equation}
 A_i(x;t) := \prod_{j\neq i}\frac{tx_i-x_j}{x_i-x_j}~,
\end{equation}
is an auxiliary function%
\footnote{This is the same function that appears in Chapter VI of Macdonald's book \cite{Macdonald:book}.}
that satisfies
\begin{equation}
 {\Delta_{q,t}(x)^{-1}}{\shift{q}{x_i} \Delta_{q,t}(x)}
 = t^{-(N-1)}A_i(x;t)\left[\shift{q}{x_i}A_i(x;t^{-1})\right]^{-1}
\end{equation}
The \(q\)-Virasoro constraint is obtained by inserting the $q$-difference operator $\oper_m$ inside the integral in \eqref{eq:partition-function-rCS}. A generalization of Stokes' theorem then states that the integral of a total $q$-derivative vanishes (provided there are no boundary contributions). The resulting equation can be written as
\begin{equation}
\label{eq:oper-constr-rCS}
 \frac{1}{N!}\int\mathd x^N\oper_m\left\{\Delta_{q,t}(x)\mathe^{-\sum_i V(x_i)}\Pi(x,\ts;q,t)\right\}
 = \left\langle\mathbb{D}_m^V\Pi(x,\ts;q,t)\right\rangle = 0
\end{equation}
Here we define the operator $\mathbb{D}_m^V$ by
\begin{equation}
\begin{aligned}
\label{eq:operbb-constr-rCS}
 \mathbb{D}_m^V f(x)
 &= \left(\Delta_{q,t}(x)\mathe^{-\sum_{i} V(x_i)}\right)^{-1}
 \oper_m\left(\Delta_{q,t}(x)\mathe^{-\sum_{i} V(x_i)}f(x)\right) \\
 &= \frac{1}{1-q}\left[\sum_{i=1}^N (q^{-\frac12}x_i)^m A_i(x;t^{-1})
 -qt^{-(N-1)}\sum_{i=1}^N (q^{\frac12}x_i)^m \mathe^{V(x_i)-V(qx_i)}
 A_i(x;t)\shift{q}{x_i}\right]f(x)
\end{aligned}
\end{equation}
This is a discrete family of deformations of the Macdonald operator $D^1_N=\sum_{i=1}^N A_i(x;t)\shift{q}{x_i}$ (also known as the Hamiltonian of the trigonometric Ruijsenaars-Schneider system), which depends non-trivially on the choice of potential $V(x)$. As a special case, when $V(x)$ is $q$-constant then we have
\begin{equation}
 \mathbb{D}^{V_{q\text{-const}}}_0 = \frac{t^{-(N-1)}}{1-q}
 \left[\frac{1-t^{N}}{1-t}-qD^1_N \right]
\end{equation}
However, the physically interesting case is when the potential is of the form \eqref{eq:potential-rCS} so that
\begin{equation}
\label{eq:shift-potential-rCS}
 \mathe^{V(x_i)-V(qx_i)} = r t^{N-1} q^{-\frac12}x_i^{-1}
\end{equation}
which is equivalent to the statement that the potential function $\mathe^{-V(x)}$ can be thought of as a meromorphic section of a line bundle of degree 1 over the elliptic curve $\mathbb{C}^\times/q^\mathbb{Z}$.

If the difference operator $\mathbb{D}^V_m$ acts on the kernel $\Pi(x,\ts;q,t)$ as multiplication by a symmetric function in the $\{x_i\}$, then one can use the identity \eqref{eq:adjointpowersum} to define a formal adjoint of $\mathbb{D}^V_m$. This is indeed possible for $m>0$, in which case we define the formal adjoint of $\mathbb{D}^V_m$ as the differential operator in times given by%
\footnote{For $m\leq 0$, $\mathbb{D}^V_m\Pi(x,\ts;q,t)$ is not polynomial in the $x_i$ but also contains negative powers. For this reason one cannot take the adjoint of those operators.}
\begin{equation}
\label{eq:adjoint-oper-rCS}
\begin{aligned}
 &(\mathbb{D}^V_m)^{\perp} =
 \oint_{|z|=1}\frac{\mathd z}{2\pi\mathi z} z^{-m}\frac{1}{(1-q)(1-t^{-1})}\Bigg\{
 1-t^{-N}\exp\left(\sum_{n=1}^\infty \frac{(q^{-\frac12}z)^n}{n}(1-t^n)t_n^\perp\right)\\
 &+prz
 \left[1-t^N\exp\left(-\sum_{n=1}^\infty\frac{(q^{\frac12}z)^{-n}}{n}(1-t^n)t_n\right)
 \exp\left(\sum_{n=1}^\infty\frac{(q^{\frac12}z)^{n}}{n}(1-t^{-n})t_n^\perp\right)
 \right]
 \Bigg\}
\end{aligned}
\end{equation}
The $q$-Virasoro constraints then can be rewritten as a set of linear differential equations for the generating function $Z^\mathrm{rCS}_N(\ts;q,t)$, namely
\begin{equation}
\label{eq:oper-constr-adjoint-rCS}
 (\mathbb{D}_m^V)^\perp Z_N^{\mathrm{rCS}}(\ts;q,t) = 0
\end{equation}
for all $m>0$.
{The integrand in \eqref{eq:adjoint-oper-rCS} can be identified with the generating function of all the constraints and it is closely related to the generating current $T(z)$ of $q$-Virasoro. We refer the reader to Appendix~C of \cite{Cassia:2020uxy} for a detailed discussion on the relation between the $q$-Virasoro constraint generators $(\mathbb{D}_m^V)^\perp$ and the generators of the $q$-Virasoro algebra.}

Observe that the factor of $z$ appearing in the beginning of the second line of \eqref{eq:adjoint-oper-rCS} is due to the variation of the CS potential \eqref{eq:shift-potential-rCS} and it induces a shift in degree in the power series expansion in $z$.
More explicitly, after taking the residue in $z$, the operator $\mathbb{D}^V_m$ becomes the sum of two pieces: the first is a differential operator of degree $m$ while the second is of degree $m-1$ in the higher times.
This shift in degree is precisely what will allow us in the next section to establish a recursion on the correlators $c_\lambda$.

\subsection{Recursive solution}
\label{sec:recursive-solution}

In this section we set out to solve the constraints in \eqref{eq:oper-constr-adjoint-rCS}. The first step is to evaluate explicitly the contour integral in \eqref{eq:adjoint-oper-rCS},
\begin{equation}
\label{eq:expand-constraint-rCS}
\begin{aligned}
 (\mathbb{D}_m^V)^\perp = -t^{-N}q^{-\frac{m}{2}}h_{m}&\left(\left\{p_n=(1-t^{n})t_n^\perp\right\}\right) 
 + pr\left(-pq^{-\frac12}\right)^{m-1}
 \Bigg[\delta_{m,1} + \\
 &-\sum_{\ell=0}^{\infty} t^{N-\ell}
 e_{\ell}\left(\left\{p_n=(1-t^n)t_n\right\}\right)
 e_{\ell+m-1}\left(\left\{p_n=(1-t^{n})t_n^\perp\right\}\right)
 \Bigg]
\end{aligned}
\end{equation}
where $h_m$ and $e_m$ are the complete homogeneous and elementary symmetric functions of degree $m$, respectively, which we think of as polynomials in the power-sums $p_n$ as in \eqref{eq:homogeneous-elementary}.

Assuming the series expansion of $Z_N^\mathrm{rCS}(\ts;q,t)$ as in \eqref{eq:generating-function-rCS}, we can let all the differential operators $t^\perp_n$ act on the monomials in times and the constraint becomes a linear system for the coefficients $c_\lambda$.
In particular this system is triangular w.r.t. a certain ordering on the partitions and the solution is unique up to overall normalization specified by the initial data $c_{\emptyset} := \langle 1\rangle = Z_N^{\mathrm{rCS}}(0;q,t)$, i.e. the empty correlator.
The proof goes along the lines of the one described in \cite{Cassia:2020uxy}.

Solving the recursion explicitly in degree 1 and 2 we get the following values for the correlators
\begin{equation}
\begin{aligned}
c_{[2]} &= \frac{ \left(t^{N-1} (q t+q+1)-1\right)}{1+t}
 \frac{1-t^N}{1-t} (rq^{\frac32}t^{N-1})^2 \, c_{\emptyset} \\
c_{[1,1]} &= \frac{\left(1-t^{N-1} (q (t-1)+1)\right)}{1-t}
 \frac{1-t^N}{1-t} (rq^{\frac32}t^{N-1})^2 \, c_{\emptyset} \\
c_{[1]} &= \frac{1-t^N}{1-t}(rq^{\frac32}t^{N-1}) \, c_{\emptyset}
\end{aligned}
\end{equation}
Similarly one can plug \eqref{eq:expand-constraint-rCS} into a computer algebra system and compute all correlators $c_\lambda$ up to arbitrarily high degree of the partition $\lambda$. While this solution is fully explicit and straightforward to compute, it is not very illuminating. In section \ref{sec:macdonald-polynomials} we discuss a more systematic way to express the solution to the constraints.

\subsection{Averages of Macdonald polynomials}
\label{sec:macdonald-polynomials}

By using the explicit solution derived in \ref{sec:recursive-solution} we can now ask what is the expectation value of a Macdonald polynomial in representation $R_\lambda$.
Using the fact that Macdonald symmetric functions admit an expansion into power-sums we can easily rewrite any Macdonald expectation value as a linear combination of the correlators $c_\lambda$ that we already computed. Computer experiments then suggest that the following formula should hold
\begin{equation}
\label{eq:macdonald-rCS}
 \frac{\langle\mcdp_\lambda(x) \rangle}{\langle 1 \rangle}
 =
 \frac{\mcdp_\lambda\left(\left\{p_n=\frac{-(-pq^{\frac12}rt^{N})^n}{1-t^n}\right\}\right)}
 {\mcdp_\lambda\left(\left\{p_n=\frac{1}{1-t^n}\right\}\right)}
 {\mcdp_\lambda\left(\left\{p_n=\frac{1-t^{nN}}{1-t^{n}}\right\}\right)}
\end{equation}
which we checked for all partitions up to $|\lambda|\leq 6$.
This is a special case of a property of certain matrix models called \textit{superintegrability} first observed in \cite{Mironov:2017och,Mironov:2017aqv,Mironov:2018ekq,Natanzon:2014mda}. Superintegrability for $q,t$-deformed matrix models was later conjectured in \cite{Morozov:2018eiq} and in \cite{Cassia:2020uxy} its was explicitly checked (up to some finite order) using the solution of the corresponding $q$-Virasoro constraints.

The last factor on the right of \eqref{eq:macdonald-rCS} is the well-known Macdonald dimension of the representation $R_\lambda$,
\begin{equation}
 \dim_{q,t}(R_\lambda)
 = \mcdp_\lambda\left(\left\{p_n=\frac{1-t^{nN}}{1-t^{n}}\right\}\right)
 = t^{\frac12|\lambda|(N-1)}\mcdp_\lambda\left(x_i=t^{\rho_i}\right)
\end{equation}
where
\begin{equation}
 \rho_i = \frac12(N-2i+1)
\end{equation}
is the Weyl vector of $U(N)$ and we can write
\begin{equation}
\label{eq:q-dimension}
 \mcdp_\lambda\left(t^{\rho}\right)
 = t^{-\lambda\cdot\rho}\prod_{\alpha>0}
 \frac{(t^{\alpha\cdot\rho+1};q)_{\alpha\cdot\lambda}}
 {(t^{\alpha\cdot\rho};q)_{\alpha\cdot\lambda}}
 = \prod_{\alpha>0} \prod_{m=0}^{{\beta}-1}
 \frac{\sin\left(\frac{\pi(m+\alpha\cdot(\beta\rho+\lambda))}{k+\beta N}\right)}
 {\sin\left(\frac{\pi(m+\alpha\cdot(\beta\rho))}{k+\beta N}\right)}
\end{equation}
where $\alpha>0$ are the positive roots of $U(N)$ {and $\alpha\cdot\lambda$ is the inner product defined in \eqref{eq:scalarprod}. Observe that the second equality in \eqref{eq:q-dimension} only makes sense for $\beta\in\mathbb{N}$.}

The fractional term on the right of \eqref{eq:macdonald-rCS} is instead a power of $q$ which for $q$ a root of unity is just a complex phase. More precisely, a direct computation shows that \eqref{eq:macdonald-rCS} becomes
\begin{equation}
 \frac{\langle\mcdp_\lambda(x) \rangle}{\langle 1 \rangle}
 = \mcdp_\lambda\left(t^{\rho}\right)
 \left[q^{\frac12\lambda\cdot\lambda} t^{\lambda\cdot\rho}
 (t^{N}rp)^{|\lambda|}\right]
\end{equation}
Fixing the FI parameter as $r=t^{-N}p^{-1}$ we precisely obtain that the overall phase corresponds to a framing factor of 1 unit
\begin{equation}
 \frac{T_{\lambda}}{T_{0}} = q^{\frac12\lambda\cdot\lambda}t^{\lambda\cdot\rho}
\end{equation}
where $T_\lambda$ is the eigenvalue of the modular $T$-matrix corresponding to the representation $R_\lambda$.

Fixing instead $r=t^{-N}q^{-1}$ we get the identity
\begin{equation}
\begin{aligned}
 \mcdp_\lambda\left(t^{\rho}\right)
 \left[q^{\frac12\lambda\cdot\lambda} t^{\lambda\cdot\rho- |\lambda|}\right]
 &= \mcdp_\lambda\left(t^{\staircase}\right)
 \left[q^{\frac12\lambda\cdot\lambda}
 t^{-\frac{|\lambda|}{2}-\frac12\lambda'\cdot\lambda'}
 \right]
\end{aligned}
\end{equation}
where the r.h.s. now matches with the result of \cite{Aganagic:2011sg} ($\staircase$ is the staircase partition of \eqref{eq:staircase}).

Because Macdonald polynomials are a complete basis of the ring of symmetric functions in the $\{x_i\}$, we can use formula \eqref{eq:macdonald-rCS} to compute expectation values of all possible observables of the model. In particular, we can give a full solution for the generating function by expanding in Macdonald polynomials (as opposed to power-sum polynomials) as
\begin{equation}
\label{eq:character-expansion}
 Z_N^{\mathrm{rCS}}(\ts;q,t)
 = \left\langle\Pi(x,\ts;q,t)\right\rangle
 = \sum_{\lambda} \left\langle\mcdp_\lambda(x)\right\rangle
 \mcdq_\lambda(\ts)
\end{equation}
{where we used the Cauchy--Littlewood identity \eqref{eq:Cauchy-Littlewood}.}
This is usually referred to as \textit{character expansion} of the generating function.

{
As a last application of the superintegrability formula \eqref{eq:macdonald-rCS} we compute a closed formula for the generating function of averages of symmetric Macdonald polynomials. If we set $t_n=y^n$ for some formal variable $y$, we obtain that the kernel $\Pi(x,\ts;q,t)$ reduces to the generating function of symmetric Macdonald polynomials
\begin{equation}
 \Pi(x,y;q,t) := \Pi(x,t_n=y^n;q,t) = \sum_{n\geq0} \mcdp_{[n]}(x) \frac{(t;q)_n}{(q;q)_n}y^n
\end{equation}
therefore
\begin{equation}
\label{eq:symmetricMacdonalds}
 \frac{Z_N^{\mathrm{rCS}}(y;q,t)}{Z_N^{\mathrm{rCS}}(0;q,t)}
 = \frac{\left\langle\Pi(x,y;q,t)\right\rangle}{\left\langle1\right\rangle}
 = \sum_{n\geq0} (prt^N q^\frac12 y)^n \frac{(t^N;q)_n}
 {(q;q)_n} q^{\binom{n}{2}}
 = \frac{(t^N;q)_\infty}{(t^N T_{q,y};q)_\infty}(-prt^N q^\frac12 y;q)_\infty
\end{equation}
which is the unique solution (up to normalization) to the $q$-difference equation
\begin{equation}
\label{eq:quantumcurve}
 (1-T_{q,y})Z_N^{\mathrm{rCS}}(y;q,t) =
 (prt^N q^{\frac12}y){(1-t^N T_{q,y})T_{q,y}}Z_N^{\mathrm{rCS}}(y;q,t)
\end{equation}
Observe that the previous equation at $r=p^{-1}t^{-N}$ (i.e. framing 1) matches the quantum mirror curve equation of the resolved conifold (see \cite{Gukov:2011qp,Bouchard:2011ya})
\begin{equation}
 \widetilde{H}_{-f}(X,Y) = 1+Y+XY^f+QXY^{f+1}
\end{equation}
for
\begin{equation}
 Q=t^N~,
 \quad\quad\quad
 X = q^{\frac12}y~,
 \quad\quad\quad
 Y = -T_{q,y}~,
 \quad\quad\quad
 f = 1
\end{equation}
which is expected by arguments of large $N$ dualities with Gromov-Witten theory. Because the equation \eqref{eq:quantumcurve} is a corollary of the $q$-Virasoro constraints, this seems to suggest an interpretation of the constraints in terms of topological recursion formulas as well.
}

\subsection{Symmetries of the constraints}
\label{sec:symmetries-rCS}

We now consider the operators \eqref{eq:adjoint-oper-rCS} and study their symmetries which will imply symmetries of the solution.

\subsubsection{Langlands duality}

It is well known that $q,t$-deformed models enjoy a non-perturbative symmetry associated to the exchange of $q$ and $t^{-1}$ which is known as ``quantum $q$-geometric'' Langlands duality \cite{Frenkel:1997,Frenkel:2010wd,Aganagic:2017smx}.
To show that our $q$-Virasoro constraints satisfy this symmetry we define the involution $\langlands$ as
\begin{equation}
\begin{array}{ccccc}
 \langlands(\beta) = \tfrac{1}{\beta}
 &\quad&
 \langlands(N) = -\beta N
 &\quad&
 \langlands(\FI) = -\frac{1}{\beta}\FI
\end{array}
\end{equation}
and one can check that $(\langlands)^2=\id$. We can extend the action of the involution to the full generating function by imposing that the times transform as well, with the rule
\begin{equation}
 \langlands(t_n) = -\frac{1-t^n}{1-q^n}p^{\frac{n}{2}}t_n
\end{equation}
so that the $q$-Virasoro operators become invariant under Langlands duality,
\begin{equation}
\label{eq:langlands-constr-rCS}
 \langlands\left((\mathbb{D}^V_m)^\perp\right)
 = (\mathbb{D}^V_m)^\perp
\end{equation}
This implies that the normalized solution to the constraints must also be invariant under the duality%
\footnote{Observe that this is not true for the partition function, in fact
\begin{equation*}
 \langlands(Z^\mathrm{rCS}_{N}(0;q,t))\neq Z^\mathrm{rCS}_{N}(0;q,t)
\end{equation*}
}. More precisely, we can write
\begin{equation}
\label{eq:langlands-generating-funtion-rCS}
 \frac{Z_N^{\mathrm{rCS}}(\{t_n\};q,t)}{Z_N^{\mathrm{rCS}}(\{0\};q,t)}
 = \frac{Z^{\mathrm{rCS}}_{-\beta N}(\{-\frac{1-t^n}{1-q^n}p^{\frac{n}{2}}t_n\};t^{-1},q^{-1})}
 {Z^{\mathrm{rCS}}_{-\beta N}(\{0\};t^{-1},q^{-1})}
\end{equation}
Observe that for arbitrary complex values of $\beta$, the dual theory has rank $-\beta N$ which generically is not a positive integer. Clearly the matrix model of this theory is not well-defined since the rank should correspond to the number of eigenvalues $x_i$. Nevertheless, the $q$-Virasoro constraints are well-defined for any complex value of the rank, hence the generating function in the r.h.s. of \eqref{eq:langlands-generating-funtion-rCS} does make sense as a formal power series solution to the constraints. In other words, one can compute the correlators for integer values of the rank and then analytically continue their expression over the whole complex plane.

\begin{remark}
As a special case of the identity \eqref{eq:langlands-generating-funtion-rCS}, we can choose the $\beta$-parameter in such a way that the rank of the model on the right becomes 1. Namely, we can fix the
value of \(\beta\) as
\begin{equation}
 \beta=-\frac{1}{N}
\end{equation}
so that we obtain the ``universal'' relation
\begin{equation}
 \frac{Z_N^{\mathrm{rCS}}(\{t_n\};t^{-N},t)}
 {Z_N^{\mathrm{rCS}}(\{0\};t^{-N},t)}
 = \frac{Z_1^{\mathrm{rCS}}(\{\frac{1-t^n}{1-t^{nN}} t^{\frac{n}{2}(N-1)}t_n\};t^{-1},t^N)}{Z_1^{\mathrm{rCS}}(\{0\};t^{-1},t^N)}
\end{equation}
Moreover, in rank \(N=1\) the solution is particularly simple,
\begin{equation}
 Z_1^{\mathrm{rCS}}(\ts;q,t)
 = \left\langle\exp\left(\sum_{n=1}^\infty \frac{x^n}{n}\frac{1-t^n}{1-q^n}t_n\right)\right\rangle
 = \sum_{m=0}^\infty \langle x^m\rangle
 h_m\left(\left\{p_n=\frac{1-t^n}{1-q^n}t_n\right\}\right)
\end{equation}
where we can evaluate the momentum $\langle x^m\rangle$ explicitly by the integral
\begin{equation}
 \langle x^m\rangle = \int_{0}^\infty\mathrm{d}x\,
 x^{m+\FI} \exp\left(-\frac{\log^2 x}{2\log q}\right)
 = q^{\frac12 (m+\FI +1)^2} \sqrt{2 \pi \log q}
\end{equation}
\end{remark}

\subsubsection{Duality at root of unity}

Another interpretation of the symmetry of the operators $(\mathbb{D}^V_m)^\perp$ comes from restricting to the physical region of parameters where $q,t$ are roots of unity as defined in \eqref{eq:qt-level}. It is known that the $q$-Virasoro algebra at root of unity degenerates to a simpler algebra related to parafermions \cite{Bouwknegt:1998,Nigro:2012av,Itoyama:2013mca,Itoyama:2014pca}. We do not pursue this avenue here however we remark that in this region of parameters the Langlands duality reduces to 3d Seiberg duality for pure CS theories, which also coincides with level-rank duality (see \cite{Kapustin:2013hpk} for a discussion of dualities in unrefined CS). The identification holds provided we assume the identity\footnote{{This constraint on $q$ and $t$ is reminiscent of the so-called \textit{wheel condition} of \cite{feigin2002symmetric}, however it is not clear to us how exactly the two concepts are related.}}
\begin{equation}
\label{eq:level-rank}
 q^k t^N = 1
\end{equation}
which follows from \eqref{eq:qt-level}. For this choice of parameters we have
\begin{equation}
 q^{k} = q^{-\beta N}
\end{equation}
so that formally we can modify the involution to act as
\begin{equation}
 \langlands(N) = k
\end{equation}
therefore, Langlands duality maps refined CS theories as
\begin{equation}
\begin{array}{ccc}
 U(N)_k &\leftrightarrow& U(k)_N \\
 &&\\
 Z^\mathrm{rCS}_{N}(q,t) &\simeq& Z^\mathrm{rCS}_{k}(t^{-1},q^{-1})
\end{array}
\end{equation}

From \eqref{eq:level-rank} it also follows that rCS theory has an additional discrete $\mathbb{Z}$-symmetry corresponding to shifts of the rank $N$ of the form
\begin{equation}
 N\mapsto N+\delta N
\end{equation}
with
\begin{equation}
 \delta N \in \tfrac{1}{\beta}\left(k+\beta N\right) \mathbb{Z}
\end{equation}
It would be interesting to find a physical motivation that might explain this discrete symmetry, perhaps via some string/M-theory construction.

\subsection{Harer-Zagier formulas}

Matrix models that possess the superintegrability property described in Section~\ref{sec:macdonald-polynomials} often have simple formulas that express the expectation value of single trace operators. In our language, these are the correlators whose partition is of length one, i.e. those of the form $c_{[n]}=\langle p_n\rangle$. Knowledge of a closed formula for such correlators can be very important because they are the coefficients of the series expansion of the 1-point resolvent of the model. It is believed that in models that exhibit superintegrability the (Laplace transforms of) single-trace correlators $c_{[n]}$ are rational functions with zeroes and poles at powers of $q$ \cite{Morozov:2020awm,Morozov:2021zmz}, and these formulas are usually called Harer-Zagier (HZ) formulas \cite{Harer:1986}.

While giving an analytic derivation of such HZ formulas can be quite complicated, guessing the actual form of the result can be done more easily. In this section we conjecture a formula of HZ-type for the CS matrix model when the refinement parameter 
$\beta$ is set to 1 (i.e. $t=q$).
First we conjecture the following closed formula for 1-point correlators (see also {Proposition 1.1} of \cite{Forrester:2020ebj})
\begin{equation}
\label{eq:HZF}
 c_{[n]} = -(-q^{N+\frac12} r)^n \frac{1-q^N}{1-q^n} \,
 \littleQJacobi^{(0,1)}_{n-1}(q^{N-1};q)
 \,c_\emptyset
\end{equation}
where $\littleQJacobi_n^{(\alpha,\beta)}(z;q)$ is a \textit{little $q$-Jacobi polynomial} defined as
\begin{equation}
\label{eq:littleQJacobi}
 \littleQJacobi^{(\alpha,\beta)}_n(z;q)
 = {}_2\phi_1(q^{-n},q^{\alpha+\beta+n+1};q^{\alpha+1};q,qz)
 = \sum_{m=0}^n
 \frac{(q^{-n};q)_m(q^{\alpha+\beta+n+1};q)_m}{(q^{\alpha+1};q)_m(q;q)_m}(qz)^m
\end{equation}
{The appearance of basic hypergeometric series is not surprising as we already noticed that symmetric correlators do resum into a generalized basic hypergeometric series as in \eqref{eq:symmetricMacdonalds}.}

It is then useful to consider the discrete Laplace transform of this formula w.r.t. the rank $N$. The transform takes the form of a generating function%
\footnote{One should regard this formula as a special case of the HZ formula for torus knots in Section 6 of \cite{Morozov:2021zmz}.}
\begin{equation}
 \sum_{N=0}^\infty \lambda^N c_{[n]} = r^n q^{\frac{n(n+2)}{2}}
 \frac{\lambda}{1-\lambda} \frac{(\lambda;q)_{n}^2}{(\lambda;q)_{2n+1}}
 \,c_\emptyset
\end{equation}
where now we can observe that the r.h.s. is indeed a rational function of $\lambda$ with zeroes and poles at integer powers of $q$. It would be interesting to derive a similar formula for the refined correlators at $t\neq q$.

\section{Refined ABJ matrix model}
\label{sec:rABJ}

The ABJ matrix model of \cite{Aharony:2008gk,Drukker:2009hy,Eynard:2014rba} admits a similar refinement deformation as that of the CS matrix model. The partition function of ABJ theory can be expressed as a matrix integral over two sets of independent variables $\{x_i\}_{i=1}^N$ and $\{y_a\}_{a=1}^M$ which can be interpreted as the eigenvalues of an Hermitean super-matrix in the Lie super-algebra of $U(N|M)$. The usual Vandermonde determinant in the measure is first substituted by the Cauchy determinant and then $q,t$-deformed to the function (see \cite{Nieri:2017ntx,Atai:2021qqi,Kimura:2021ngu})
\begin{equation}
 \Delta_{q,t}(x,y) = \frac{\Delta_{q,t}(x)\Delta_{t^{-1},q^{-1}}(y)}
 {\prod_{i=1}^N\prod_{a=1}^M (1-\sqrt{t/q}~x_i/y_a)(1-\sqrt{t/q}~y_a/x_i)}
\end{equation}
The potential of the model is also modified in order to account for the second set of $\{y_a\}$ variables and we write it as
\begin{equation}
\begin{aligned}
 V(x,y) =&
  \sum_{i=1}^N\frac{\log^2x_i}{2\log q}
 + \big(M-\beta(N-1)-\FI\big)\sum_{i=1}^N\log x_i \\
 & + \sum_{a=1}^M\frac{\log^2y_a}{2\log t^{-1}}
 + \big(N-\tfrac{1}{\beta}(M-1)+\tfrac{1}{\beta}\FI\big)\sum_{a=1}^M\log y_a
\end{aligned}
\end{equation}
so that the first set of eigenvalues $x_i$ has coupling $\log q$ while the eigenvalues $y_a$ have coupling $\log t^{-1}$ (the FI parameters are also related in such a way as to preserve symmetry under Langlands duality). Observe that this is the refinement deformation of the usual statement that, if we regard ABJ theory as a quiver gauge theory with two nodes $U(N)$ and $U(M)$, then the two CS levels must be opposite to each other\footnote{In the case of the ABJ theory, supersymmetry prevents the CS level to receive quantum corrections from the dual Coxeter number. Therefore one has $q=\mathe^{{2\pi\mathi}/{k}}$ instead of \eqref{eq:qt-level}. A similar conclusion was reached in \cite{Mikhaylov:2014aoa} in the case of super-group CS theory.}.

Expectation values of gauge invariant operators are given by averages of symmetric functions in the eigenvalues $\{x_i\}$ and $\{y_a\}$. In the following we restrict ourselves to those symmetric functions which can be written as polynomials in the deformed power-sums of \cite{Atai:2021qqi} (see also \cite{sergeev2008deformed}), defined as
\begin{equation}
\label{eq:power-sums-rABJ}
 p_n(x,y) := p_n(x)-\frac{1-q^n}{1-t^n}p^{-\frac{n}{2}}p_n(y)
\end{equation}
The reproducing kernel which gives the generating function of such deformed power-sums is the function
\begin{equation}
 \Pi(x,y,\ts;q,t) = \exp\left(\sum_{n=1}^\infty
 \frac{t_n}{n}\frac{1-t^n}{1-q^n}p_n(x,y)\right)
\end{equation}
which can be obtained from \eqref{eq:reproducing-kernel-rCS} by substituting $p_n(x)$ with $p_n(x,y)$. It follows that
\begin{equation}
\label{eq:adjoint-rABJ}
 p_n(x,y)\Pi(x,y,\ts;q,t) = t^\perp_n\Pi(x,y,\ts;q,t)
\end{equation}
Finally, we can write down the generating function of the rABJ matrix model as
\begin{equation}
\label{eq:generating-function-rABJ}
 Z_{N,M}^{\mathrm{rABJ}}(\ts;q,t) = \frac{1}{N!M!}\int_{\mathbb{R}^{N+M}_{+}}
 \Delta_{q,t}(x,y)
 \Pi(x,y,\ts;q,t)
 \mathe^{-V(x,y)}
 \prod_{i=1}^N\mathrm{d}x_i
 \prod_{a=1}^M\mathrm{d}y_a \,
\end{equation}
As already mentioned, in the unrefined limit $t=q$ the measure $\Delta_{q,t}(x,y)$ reduces to a Cauchy determinant and one ends up with the usual matrix model for the ABJ theory which is the super-group version of \eqref{eq:unrefined-CS} (see \cite{Eynard:2014rba} for details).
Moreover, it is evident from the definitions that the rABJ matrix model contains rCS as the special case $M=0$, i.e.
\begin{equation}
 \label{eq:rABJ-to-rCS}
 Z^\mathrm{rCS}_N(\ts;q,t) = Z^\mathrm{rABJ}_{N,0}(\ts;q,t)
\end{equation}

\begin{remark}
Convergence of the matrix integral for the rABJ partition function requires a careful choice of contour of integration depending on the values of the parameters $q$ and $t$. In the following we assume that such a choice of contour always exists, however we do not write down the definition of the contour explicitly as it is known that $q$-Virasoro constraints are independent of such choice. Moreover, as for the case of rCS, one could even define a discrete version of the rABJ model using the substitutions $\int\mathd x_i \mapsto \int\mathd_q x_i$ and $\int\mathd y_a \mapsto \int\mathd_{t^{-1}}y_a$, which would satisfy the same set of $q$-Virasoro constraints.

\end{remark}

\subsection{q-Virasoro constraints}

In order to derive the appropriate $q$-Virasoro constraints for the rABJ model we need to find a generalization of the difference operator $\oper_m$ in \eqref{eq:q-diff-operator}
which acts both on the $x_i$ variables and on the $y_a$ variables. The fact that these two sets of variables have different couplings $q$ and $t^{-1}$, respectively, suggests that their shifts should also be different. By using the symmetry under the exchange of $x_i\leftrightarrow y_a$ and $q\leftrightarrow t^{-1}$ we are lead to define the operators
\begin{equation}
\begin{aligned}
 \oper_m f(x,y) =&
 \sum_{i=1}^N D_{q,x_i}
 \left[q^{-\frac{m}{2}}x_i^{m+1} A_i(x,y;q^{-1},t^{-1})f(x,y)\right] \\
 & + \sum_{a=1}^M D_{t^{-1},y_a}
 \left[t^{\frac{m}{2}}y_a^{m+1} B_a(x,y;q^{-1},t^{-1})f(x,y)\right]
\end{aligned}
\end{equation}
where
\begin{equation}
 A_i(x,y;q,t) =
 \prod_{j\neq i}\frac{tx_i-x_j}{x_i-x_j}
 \prod_{a=1}^M \frac{y_a-\sqrt{t/q}x_i}{y_a-q\sqrt{t/q}x_i}
\end{equation}
\begin{equation}
 B_a(x,y;q,t) =
 \prod_{b\neq a}\frac{q^{-1}y_a-y_b}{y_a-y_b}
 \prod_{i=1}^N \frac{x_i-\sqrt{t/q}y_a}{x_i-t^{-1}\sqrt{t/q}y_a}
\end{equation}
We now show that this are the correct difference operators to derive the rABJ generalization of the $q$-Virasoro constraints. First we observe that the functions $A_i(x,y;q,t)$ and $B_a(x,y;q,t)$ satisfy
\begin{equation}
 \Delta_{q,t}(x,y)^{-1}\shift{q}{x_i}\Delta_{q,t}(x,y)
 = q^{M}t^{-(N-1)}A_i(x,y;q,t)\left[\shift{q}{x_i}A_i(x,y;q^{-1},t^{-1})\right]^{-1}
\end{equation}
and
\begin{equation}
 \Delta_{q,t}(x,y)^{-1}\shift{t^{-1}}{y_a}\Delta_{q,t}(x,y)
 = t^{-N}q^{M-1}B_a(x,y;q,t)\left[\shift{t^{-1}}{y_a}B_a(x,y;q^{-1},t^{-1})\right]^{-1}
\end{equation}
With some straightforward algebraic manipulations we can rewrite the constraints as
\begin{equation}
 \left\langle \mathbb{D}_m^V \Pi(x,y,\ts;q,t) \right\rangle = 0
\end{equation}
where
\begin{equation}
\begin{aligned}
 \mathbb{D}_m^V
 =& \frac{1}{1-q}\Bigg[\sum_{i=1}^N (q^{-\frac12}x_i)^m A_i(x,y;q^{-1},t^{-1}) \\
  & \quad\quad\quad -q^{M+1}t^{-(N-1)}\sum_{i=1}^N (q^{\frac12}x_i)^m \mathe^{V(x,y)-\shift{q}{x_i}V(x,y)}
 A_i(x,y;q,t)\shift{q}{x_i}\Bigg] \\
 & + \frac{1}{1-t^{-1}}\Bigg[\sum_{a=1}^M (t^{\frac12}y_a)^m B_a(x,y;q^{-1},t^{-1}) \\
 & \quad\quad\quad\quad\quad -t^{-(N+1)}q^{M-1}\sum_{a=1}^M (t^{-\frac12}y_a)^m \mathe^{V(x,y)-\shift{t^{-1}}{y_a}V(x,y)}
 B_a(x,y;q,t)\shift{t^{-1}}{y_a}\Bigg]
\end{aligned}
\end{equation}
which reduces to \eqref{eq:operbb-constr-rCS} when $M=0$. {One should also notice that for $m=0$ and $V(x,y)$ constant under $x_i\to qx_i$ and $y_a\to t^{-1}y_a$, the operator $\mathbb{D}_0^V$ can be written in terms of the deformed Macdonald-Ruijsenaars operator of \cite{sergeev2004deformed}.}

The last step of the derivation consists in computing the formal adjoint of the operators $\mathbb{D}_m^V$ using \eqref{eq:adjoint-rABJ}, and we find
\begin{equation}
\label{eq:adjoint-oper-rABJ}
\begin{aligned}
 &(\mathbb{D}^V_m)^{\perp} =
 \oint_{|z|=1}\frac{\mathd z}{2\pi\mathi z} z^{-m}\frac{1}{(1-q)(1-t^{-1})}\Bigg\{
 1-q^{M}t^{-N}\exp\left(\sum_{n=1}^\infty \frac{(zq^{-\frac12})^n}{n}(1-t^n)t_n^\perp\right)\\
 &+prz
 \left[1-q^{-M}t^N\exp\left(-\sum_{n=1}^\infty\frac{(q^{\frac12}z)^{-n}}{n}(1-t^n)t_n\right)
 \exp\left(\sum_{n=1}^\infty\frac{(q^{\frac12}z)^{n}}{n}(1-t^{-n})t_n^\perp\right)
 \right]
 \Bigg\}
\end{aligned}
\end{equation}
so that the constraints become
\begin{equation}
 (\mathbb{D}_m^V)^\perp Z^\mathrm{rABJ}_{N,M}(\ts;q,t) = 0~,
 \quad\quad\quad
 m\geq1
\end{equation}

A couple of observations are in order. First we remark that the operator in \eqref{eq:adjoint-oper-rABJ} can formally be obtained from \eqref{eq:adjoint-oper-rCS} via the substitution
\begin{equation}
\label{eq:effective-rank-def}
 N \mapsto N_\mathrm{eff} := N-\tfrac{1}{\beta}M
\end{equation}
so that we can use the $q$-Virasoro constraints to argue that there exists an equivalence
between rABJ theory for the super-group $U(N|M)$ and rCS for $U(N-\frac{1}{\beta}M)$.
The precise statement of this equivalence is given by the identity
\begin{equation}
\label{eq:effective-rank}
 \frac{Z_{N,M}^\mathrm{rABJ}(\ts;q,t)}{Z_{N,M}^\mathrm{rABJ}(0;q,t)}
 = \frac{Z_{N-\frac{1}{\beta}M}^\mathrm{rCS}(\ts;q,t)}{Z_{N-\frac{1}{\beta}M}^\mathrm{rCS}(0;q,t)}
\end{equation}
between normalized rCS and rABJ generating functions. Because the $q$-Virasoro constraints in rCS are completely solvable (up to overall normalization), it follows that the rABJ generating function is also uniquely defined by $q$-Virasoro.
{Moreover, because the operators \eqref{eq:adjoint-oper-rABJ} and \eqref{eq:adjoint-oper-rCS} coincide up to redefinition of the rank, it follows that they satisfy the same algebraic relations, i.e. those given by the $q$-Virasoro algebra.}

As a second observation we can ask the following question: given some specific values of
\(\beta\) and \(N_{\mathrm{eff}}\), are there two integers \(N,M\) such
that \(N_\mathrm{eff}=N-\frac{1}{\beta}M\)? If this is the case, then we can use the correspondence in \eqref{eq:effective-rank} to define rCS theory for non-integer rank $N_{\mathrm{eff}}$ (which does not make sense as an honest matrix model) by instead computing the generating function of rABJ theory with integer ranks $N,M$ as in the r.h.s. of \eqref{eq:effective-rank-def}.

We will now provide a full answer to this question in the case that \(\beta\in\mathbb{Q}\). Let \(\beta=-\frac{\epsilon_2}{\epsilon_1}\) where $\epsilon_1,\epsilon_2\in\mathbb{Z}$ and we can assume without loss of generality that \(\gcd(\epsilon_1,\epsilon_2)=1\). We want to find
\(N,M\in\mathbb{Z}\) such that
\begin{equation}
 N_\mathrm{eff}=N+\frac{\epsilon_1}{\epsilon_2}M
\end{equation}
If \(N_\mathrm{eff}\in\mathbb{Z}\) we can just take
\(N=N_\mathrm{eff}\) and \(M=0\). If \(N_\mathrm{eff}\) is not integer
then the previous equation implies that it must be a rational number. We
can then assume that \(N_\mathrm{eff}=\frac{a}{b}\) for some coprime
integers \(a,b\). With these assumptions the equation we need to solve
becomes
\begin{equation}
 b \epsilon_2 N + b \epsilon_1 M = a \epsilon_2
\end{equation}
This linear Diophantine equation admits an integer
solution iff there exists an integer \(e\) such that
\begin{equation}
 e b = a \epsilon_2
 \quad\quad\Leftrightarrow\quad\quad
 N_\mathrm{eff} = \frac{a}{b} = \frac{e}{\epsilon_2}
\end{equation}
If \(N_\mathrm{eff}\) has this specific form, then we can
use Bézout's lemma to find two integers $n,m$ that solve the equation
\begin{equation}
 \epsilon_2 n + \epsilon_1 m = 1
\end{equation}
It then follows that
\begin{equation}
 N_\mathrm{eff} = \frac{e}{\epsilon_2} = e\frac{\epsilon_2 n + \epsilon_1 m}{\epsilon_2}
 = (en)-\frac{1}{\beta}(em)
\end{equation}
therefore we can take \(N=en\) and \(M=em\) as integer solutions to the problem.

As a final remark, we observe that the choice of rational $\beta$ has a physical interpretation in terms of Virasoro minimal models. More precisely, the Virasoro central charge of a minimal model is parametrized by two integers $\epsilon_1,\epsilon_2$ as
\begin{equation}
 c_{\epsilon_1,\epsilon_2} = 1+6\frac{(\epsilon_1+\epsilon_2)^2}{\epsilon_1\epsilon_2}
 = 1-6\left(\sqrt{\beta}-\frac{1}{\sqrt{\beta}}\right)^2
\end{equation}
This suggests that there might be a connection between rABJ theories at rational $\beta$ and Virasoro minimal models, or rather their $q$-deformations, which could be explained by the BPS/CFT correspondence.

\subsection{Langlands duality}
\label{langlands-duality}

The ``quantum $q$-geometric'' Langlands duality observed for the rCS matrix model generalizes\footnote{{To the best of our knowledge there is no interpretation of Langlands duality for super-CS theories in terms of level-rank duality.}} to the case of refined ABJ, where the action of the involution $\langlands$ is much more obvious in that it now exchanges the degrees of freedom $\{x_i\}$ with the $\{y_a\}$ while also exchanging $q\leftrightarrow t^{-1}$ as well as $N\leftrightarrow M$.
More specifically, by using
\begin{equation}
\begin{aligned}
 \langlands(t_n) &= -\frac{1-t^n}{1-q^n}p^{\frac{n}{2}} t_n \\
 \langlands(p_n(x,y)) &= -\frac{1-t^n}{1-q^n}p^{\frac{n}{2}} p_n(x,y)
\end{aligned}
\end{equation}
we observe that not just the operators $\mathbb{D}^V_m$ and its adjoint are manifestly invariant under $\langlands$ but also the actual generating function in \eqref{eq:generating-function-rABJ},
\begin{equation}
 \langlands\left({Z^\mathrm{rABJ}_{N,M}(\ts;q,t)}\right)
 = {Z^\mathrm{rABJ}_{M,N}(\langlands(\ts);t^{-1},q^{-1})}
 = {Z^\mathrm{rABJ}_{N,M}(\ts;q,t)}
\end{equation}
where the first equality follows from the definition of the involution $\langlands$ while the second equality follows from the definition of the generating function ${Z^\mathrm{rABJ}_{N,M}(\ts;q,t)}$.
Notice that this is not the case for the generating function of rCS, where the symmetry is not manifest at the level of the integral but it becomes so only when the $q$-Virasoro solution is normalized by the partition function.

Moreover, we can use this improved understanding of Langlands duality in rABJ theory to explain the action of $\langlands$ in rCS. To do so, we use the identification
\eqref{eq:rABJ-to-rCS} to which we apply $\langlands$ to obtain
\begin{equation}
 \langlands\left(\frac{Z^\mathrm{rCS}_N(\ts;q,t)}{Z^\mathrm{rCS}_N(0;q,t)}\right)
 = \langlands\left(\frac{Z^\mathrm{rABJ}_{N,0}(\ts;q,t)}{Z^\mathrm{rABJ}_{N,0}(0;q,t)}\right)
 = \frac{Z^\mathrm{rABJ}_{0,N}(\langlands(\ts);t^{-1},q^{-1})}{Z^\mathrm{rABJ}_{0,N}(0;t^{-1},q^{-1})}
 = \frac{Z^\mathrm{rCS}_{-\beta N}(\langlands(\ts);t^{-1},q^{-1})}{Z^\mathrm{rCS}_{-\beta N}(0;t^{-1},q^{-1})}
\end{equation}
where in the last equality we used \eqref{eq:effective-rank}.
This provides a better explanation for the transformation rule of the rCS matrix model.

\subsection{Averages of Super-Macdonald polynomials}

Similarly to the case of rCS, we can study superintegrability of characters for rABJ theory. If we regard rABJ as the super-group version of rCS we can in principle consider expectation values of any irreducible character of $U(N|M)$. These correspond to insertions of unknot Wilson loop in irreducible representations of the super-group and these were already considered in \cite{Mikhaylov:2014aoa,Mikhaylov:2015nsa} for the unrefined case. There it was argued that on $S^3$ the $U(N|M)$ theory can be Higgsed down to $U(N-M)$ so that the expectation value of a Wilson loop labeled by a maximally atypical representation of the super-group is equal to the expectation value
of the corresponding Wilson loop for the bosonic theory at the effective rank $N-M$.
We argue that this symmetry breaking phenomenon should admit a refinement which is the physical manifestation of the correspondence in \eqref{eq:effective-rank}.

In this paper we do not consider arbitrary irreducible representations of $U(N|M)$, but we restrict to those whose characters can be written as polynomial combinations of the deformed power-sums in \eqref{eq:power-sums-rABJ}. These characters belong to the ring of symmetric function in the variables $\{x_i\}$ and $\{y_a\}$, which is known to be generated by a basis of super-Macdonald polynomials $\supermacdonald_\lambda(x,y)$ defined in \cite{sergeev2008deformed,Atai:2021qqi}. These polynomials are obtained from the usual Macdonald functions by specializing the power-sums $p_n$ to the deformed power-sums $p_n(x,y)$, i.e.
\begin{equation}
\label{eq:superMcdP}
 \supermacdonald_\lambda(x,y)
 = \mcdp_\lambda\left(\{p_n=p_n(x,y)\}\right)
\end{equation}
Now, using that
\begin{equation}
 \langle p_n(x,y) \rangle
 = t^\perp_n Z^\mathrm{rABJ}_{N,M}(\ts,q,t)|_{\ts=0}
\end{equation}
we can immediately generalize the superintegrability formula \eqref{eq:macdonald-rCS} by substituting the rank $N$ with the effective rank $N-\tfrac{1}{\beta}M$ to give
\begin{equation}
\label{eq:macdonald-rABJ}
 \frac{\langle\supermacdonald_\lambda(x,y) \rangle}{\langle 1 \rangle}
 =
 \frac{\mcdp_\lambda\left(\left\{p_n=\frac{-(-pq^{\frac12}rt^{N}q^{-M})^n}{1-t^n}\right\}\right)}
 {\mcdp_\lambda\left(\left\{p_n=\frac{1}{1-t^n}\right\}\right)}
 {\mcdp_\lambda\left(\left\{p_n=\frac{1-t^{nN}q^{-nM}}{1-t^{n}}\right\}\right)}
\end{equation}
The rABJ generating function admits a similar character expansion to that in \eqref{eq:character-expansion}, therefore knowledge of the expectation value of Super-Macdonald polynomials implies full knowledge of the function $Z^\mathrm{rABJ}_{N,M}(\ts;q,t)$.

\section{Conclusions and discussions}
\label{sec:conclusions-and-discussions}

In this paper we considered matrix models for a refinement of CS theory and ABJ theory and we showed that there is an action of the $q$-Virasoro algebra on their generating functions of observables. Moreover, we argued that the condition that such generating functions are annihilated by all positive $q$-Virasoro generators, provides an infinite set of homogeneous constraints which admit a unique solution in the space of formal power series in higher times. We derived such solutions both by recursion on the correlation functions and also as character expansion formulas in terms of Macdonald averages. Our methods provide an alternative point of view on the results of \cite{Aganagic:2011sg,Aganagic:2012ne} for rCS while they give new interesting predictions for the less-studied refinement of ABJ theory (and ABJM when $N=M$).

Here we list some related questions and future research directions.
\begin{itemize}
\item The first and perhaps most obvious question one can ask is whether the results of this paper can be extended to the case of CS theory in the presence of matter fields. In the series of papers \cite{Lodin:2018lbz,Cassia:2019sjk,Cassia:2020uxy} the case of 3d $\mathcal{N}=2$ Yang-Mills with an adjoint matter field and $N_f$ fundamental (anti-)chirals was considered and $q$-Virasoro constraints were derived and solved. However, a strong technical requirement for the complete solvability of the equations was that the effective CS level be set to zero. In the present paper the situation is somewhat reversed in the sense that the level can be chosen arbitrarily but there are no matter fields. It is not clear to us if these are just some technical difficulties or whether there is a deeper relation between uniqueness of the solution of the constraints and the specific choice of parameters and field content of the 3d gauge theory.

\item Another natural extension of our results would be that of considering $q$-Virasoro constraints in matrix models for more general knots, such as the torus knot matrix models. Even though the approach of \cite{Aganagic:2011sg,Aganagic:2012ne} using the explicit form of the Kac-Peterson modular matrices has been successful in computing refined knot invariants for torus knots, it is still not clear how to write down a refined matrix model in that case. We leave the investigation of $q$-Virasoro for torus knots matrix models for future projects, however we do make some remarks of the relation between torus knot invariants and unknot invariants in Appendix \ref{sec:remarks-on-torus-knots}, alas, for the unrefined case.

\item In \cite{vanDiejen:2021eiw} an elliptic deformation of the fusion ring of $\widehat{su}(N)_k$ was introduced and the corresponding elliptic deformations of the modular $S$ and $T$ matrices were also computed. This deformation is known to reproduce trigonometric/Macdonald deformation when the elliptic parameter is set to zero. It would be interesting to investigate whether there exists an elliptic Virasoro \cite{Nieri:2015dts} version of the constraints for the corresponding elliptic deformation of CS matrix model and whether those constraints are uniquely solvable as in the case of the present paper.

\item Finally, a central question raised by our results is whether there exists a wider class of potentials $V(x)$ such that the $q$-Virasoro constraints can be solved fully and uniquely in terms of correlators or averages of characters. Namely, in \cite{Lodin:2018lbz,Cassia:2020uxy} it was shown that potentials of the form $V(x) = \sum_{k=1}^{N_f}\li_2(u_k x;q)$ satisfying
\begin{equation*}
 \mathe^{V(x)-V(qx)} = \prod_{k=1}^{N_f}(1-u_k x)^{-1}
\end{equation*}
lead to a full solution of the constraints for $N_f=1,2$, where $N_f$ is the step of the recursion. For higher $N_f$ it was shown that the constraints cannot fix all correlators but only a subset and that the remaining correlators do depend on additional non-canonical choices of initial data for the recursion. In this paper we instead argue that there is another potential which also leads to constraints that have a unique solution only depending on a choice of normalization. The $q$-shift transformation of the rCS potential $V(x) = \tfrac{\log^2 x}{2\log q}$ is
\begin{equation*}
 \mathe^{V(x)-V(qx)} = q^{-\frac12}x^{-1}
\end{equation*}
and it induces a recursion of step 1. It appears from this discussion that whenever
\begin{equation*}
 \mathe^{V(x)-V(qx)} = (\text{polynomial in $x$ of degree $d$})^{-1}
\end{equation*}
then the $q$-Virasoro constraints can be written as recursion relations between correlators of order $m$ and correlators of order $m-d$, which appear to be uniquely solvable only for $d=1$ and $d=2$. It is not clear to us whether these are the only cases for which the solution of the constraints is unique.
\end{itemize}

\section*{Acknowledgments}

{The authors would like to thank the anonymous referees for carefully reading the manuscript and providing many interesting suggestions.} 
All authors are supported in part by the grant ``Geometry and Physics'' from the Knut and Alice Wallenberg foundation.

\appendix

\section{Symmetric functions}
\label{sec:symmetric-functions}

For completeness we collect some basic notions on the theory of symmetric functions. See \cite{Macdonald:book} for more details {and proofs of the statements in this section}.

We start by recalling some definitions about partitions. An integer \textit{partition} $\lambda$ is a sequence of non-increasing integer numbers
\begin{equation}
 \lambda = [\lambda_1,\lambda_2,\dots]
\end{equation}
with a finite number of non-zero terms. We denote by $\ell(\lambda)$ the \textit{length} of the partition, defined as
\begin{equation}
 \ell(\lambda) = \mathrm{Card}\{j:\lambda_j\neq0\}
\end{equation}
and we denote $|\lambda|$ the \textit{weight} or \textit{size} or \textit{degree} of the partition,
\begin{equation}
 |\lambda| = \sum_{i\geq1}\lambda_i
\end{equation}
and we say that $\lambda\vdash n$ if $|\lambda|=n$.
The \textit{conjugate} partition to $\lambda$ is denoted by $\lambda'$.
For later convenience we also introduce the \textit{staircase} partition $\staircase$ defined as
\begin{equation}
\label{eq:staircase}
 \staircase = [N-1,N-2,\dots,1,0]
\end{equation}
We define $\Aut(\lambda)$ as the group of automorphisms of a given partition, and its cardinality can be computed as
\begin{equation}
 |\Aut(\lambda)| = \prod_{i\geq1} \mathrm{Card}\{j:\lambda_j=i\}!
\end{equation}
It is also useful to define the function
\begin{equation}
 z_\lambda = |\Aut(\lambda)|\prod_{i\geq1}\lambda_i
\end{equation}
Additionally we define a scalar product on partitions as
\begin{equation}
\label{eq:scalarprod}
 \lambda\cdot\mu = \sum_{i\geq1} \lambda_i\, \mu_i
\end{equation}
\textit{Symmetric functions} are elements of the ring $\Lambda$ defined as the inverse limit of the graded ring $\Lambda_N=\mathbb{Q}[x_1,\dots,x_N]^{S_N}$ of symmetric polynomials in $N$ variables with respect to the projections $\Lambda_{M}\to\Lambda_{N}$ for $M\geq N$. The ring $\Lambda$ has many bases, among which the elementary symmetric functions $e_n$, the complete homogeneous symmetric functions $h_n$, the power-sums symmetric functions $p_n$ and the Schur symmetric functions $\schur_\lambda$.
Upon defining the power-sums as
\begin{equation}
 p_n = \sum_{i\geq1}x_i^n~,
 \quad\quad\quad
 p_\lambda = \prod_{i\geq1} p_{\lambda_i}
\end{equation}
we can obtain all other basis via polynomial combinations, e.g.
\begin{equation}
\label{eq:homogeneous-elementary}
 h_m = \sum_{\lambda\vdash m} {z_\lambda^{-1}}{p_\lambda}~,
 \quad\quad\quad
 e_m = \sum_{\lambda\vdash m} (-1)^{\ell(\lambda)}{z_\lambda^{-1}}{p_\lambda}
\end{equation}
\begin{equation}
\label{eq:schur}
 \schur_\lambda = \det_{1\leq i,j\leq \ell(\lambda)}
 \left[h_{\lambda_i-i+j}\right]
\end{equation}
Moreover, there exists an inner product on symmetric functions defined by the relations
\begin{equation}
 (p_\lambda,p_\mu) = z_\lambda \delta_{\lambda\mu}
\end{equation}
or equivalently
\begin{equation}
 (\schur_\lambda,\schur_\mu) = \delta_{\lambda\mu}
\end{equation}
Let $\Lambda_{q,t}=\Lambda\otimes\mathbb{F}_{q,t}$ with $\mathbb{F}_{q,t}$ the field of rational functions in $q$ and $t$. Then there exists a deformed inner product defined by
\begin{equation}
\label{eq:z-lambda-qt}
 (p_\lambda,p_\mu)_{q,t} = z_\lambda(q,t)\delta_{\lambda\mu}~,
 \quad\quad\quad
 z_\lambda(q,t) = z_\lambda \prod_{i\geq1}
 \frac{1-q^{\lambda_i}}{1-t^{\lambda_i}}
\end{equation}
Macdonald functions are defined as the $q,t$-deformation of Schur symmetric functions which are orthogonal w.r.t. the inner product $(\cdot,\cdot)_{q,t}$. More precisely, there are two dual basis of Macdonald functions, $\mcdp_\lambda$ and $\mcdq_\mu$ with
\begin{equation}
 (\mcdp_\lambda,\mcdq_\mu)_{q,t} = \delta_{\lambda\mu}
\end{equation}
{
where $\mcdp_\lambda$ and $\mcdq_\lambda$ are related by
\begin{equation}
\label{eq:mcdPQ}
 \mcdp_\lambda \prod_{(i,j)\in\lambda} (1-q^{\lambda_i-j}t^{\lambda_j'-i+1})
 = \mcdq_\mu \prod_{(i,j)\in\lambda} (1-q^{\lambda_i-j+1}t^{\lambda_j'-i})
\end{equation}
so that they are essentially the same symmetric function up to an overall combinatorial factor of $q$ and $t$.
}

For $f\in\Lambda_{q,t}$ we use the plethystic notation $f(\{p_n\})$ to indicate that it is a symmetric function written as a polynomial in the power-sums, while we use the notation $f(x)\equiv f(x_1,x_2,\dots)$ to indicate that it is an actual symmetric polynomial in the variables $\{x_i\}$.

{
Given two sets of independent variables $\{x_i\}$ and $\{y_j\}$, one defines the Macdonald reproducing kernel
\begin{equation}
 \Pi(x,y;q,t) = \prod_{i,j} \frac{(tx_iy_j;q)_\infty}{(x_iy_j;q)_\infty}
\end{equation}
then one can rewrite the kernel in the following equivalent ways
\begin{equation}
\label{eq:Cauchy-Littlewood}
\begin{aligned}
 \Pi(x,y;q,t) &= \exp\left(\sum_{n=1}^\infty
 \frac{p_n(x)p_n(y)}{n}\frac{1-t^n}{1-q^n} \right) \\
 &= \sum_{\lambda} z_\lambda(q,t)^{-1} p_\lambda(x) p_\lambda(y) \\
 &= \sum_{\lambda} \mcdp_\lambda(x) \mcdq_\lambda(y)
\end{aligned}
\end{equation}
which is the celebrated Cauchy--Littlewood identity for Macdonald functions.
}

\section{Remarks on torus knots}
\label{sec:remarks-on-torus-knots}

In the main part of the paper we considered matrix models for unknot Wilson loops. In this section we provide some remarks for the case of torus knots. This is the only other class of knots for which a matrix model description has been derived \cite{Lawrence:1999lr,Beasley:2009mb,Kallen:2011ny,Brini:2011wi}, unfortunately no refinement of this matrix model has been worked out yet. The general expectation is that there should be a matrix model computing knot polynomials for any kind of knot, and perhaps each of such matrix model should admit a refinement.

It would be interesting to ask whether there exists a $q$-Virasoro action on the generating function of Wilson loop expectation values in all of these matrix models\footnote{See \cite{Dubinkin:2013tda} for earlier attempts of using classical Virasoro constraints to compute correlators in unrefined torus knot matrix models.}. For the case of torus knots matrix models the $q$-difference operator in \eqref{eq:q-diff-operator} at $t=q$ does not seem to give rise to recursion equations that can be solved uniquely. A more subtle modification of that operator is needed and we leave that for future investigations. Here we make some observations on how to employ the solution to $q$-Virasoro constraints derived in section \ref{sec:q-virasoro-constraints} to make predictions about torus knots expectation values.

The $(P,Q)$-torus knot generating function is
\begin{equation}
\label{eq:torus-knots}
\begin{aligned}
 Z^{(P,Q)}_{N}(\ts;q) = \frac{(PQ)^N}{N!}\int_{\mathbb{R}^N_{+}}
 \Delta(x^P)\Delta(x^Q)
 \prod_{i=1}^N \exp\left[-PQ\frac{\log^2x_i}{2\log{q}}
 +\FI\log x_i + \sum_{n=1}^\infty \frac{x_i^{nPQ} t_n}{n}\right]\mathrm{d}x_i
\end{aligned}
\end{equation}
where \(P,Q\) are positive integers defining the knot and
\begin{equation}
 \Delta(x) = \prod_{i< j} (x_i-x_j) = \det_{1\leq i,j\leq N} [x_i^{\staircase_j}]
\end{equation}
is the Vandermonde determinant (with $\staircase$ as in \eqref{eq:staircase}). Using the identity
\begin{equation}
\label{eq:weyl-character}
 \frac{\Delta(x^Y)}{\Delta(x)}
 = \frac{\det_{1\leq i,j\leq N}[x_i^{\staircase_j+(Y-1)\staircase_j}]}
 {\det_{1\leq i,j\leq N}[x_i^{\staircase_j}]}
 = \schur_{(Y-1)\staircase}(x)~,
 \quad\quad\quad
 Y=P,Q
\end{equation}
which follows from Weyl's character formula for Schur polynomials, we can write
\begin{equation}
\label{eq:CS-to-torusCS}
\begin{aligned}
 Z^{(P,Q)}_{N}(\ts;q) = & \frac{(PQ)^N}{N!}\int_{\mathbb{R}^N_{+}}
 \prod_{i=1}^N\mathrm{d}x_i\,
 |\Delta(x)|^2 \schur_{(P-1)\staircase}(x) \schur_{(Q-1)\staircase}(x) \times\\
 & \times \prod_{i=1}^N \exp\left[-PQ\frac{\log^2x_i}{2\log{q}}
 +\FI\log x_i + \sum_{n=1}^\infty \frac{x_i^{nPQ} t_n}{n}\right] \\
 = & (PQ)^N \schur_{(P-1)\staircase}(\{u_n^\perp\})
 \schur_{(Q-1)\staircase}(\{u_n^\perp\})
 \left.Z_N^{\mathrm{CS}}({\bf u};q^{\frac{1}{PQ}})\right|_{u_n=PQ\delta_{n|PQ}t_{n/PQ}}
\end{aligned}
\end{equation}
where in the last line we rewrote the torus knot generating function as the action of two adjoint Schur polynomials on the generating function of (unrefined) CS matrix model at the coupling $q^{\frac{1}{PQ}}$. The times of the two generating functions are identified using the fact that
\begin{equation}
 \sum_{n=1}^\infty \frac{x_i^{nPQ} t_n}{n}
 = \sum_{n=1}^\infty \frac{x_i^n}{n} (PQ\delta_{n|PQ}t_{{n}/{PQ}})
\end{equation}
where $\delta_{n|PQ}$ is 1 if $n$ is a multiple of $PQ$ and 0 otherwise.

Since $q$-Virasoro constraints give a full solution for $Z_N^{\mathrm{CS}}({\bf u};q^{\frac{1}{PQ}})$, then we also have a combinatorially explicit way to compute $Z^{(P,Q)}_{N}(\ts;q)$ just by acting with an appropriate differential operator of finite degree in times.

While it is tempting to conjecture the form of refined CS matrix model for torus knots by deforming Schur polynomials in \eqref{eq:CS-to-torusCS} to Macdonald polynomials, we must observe that the identity \eqref{eq:weyl-character} does not generalize to the Macdonald level and therefore some other techniques are required in order to treat the refined model.

\section{{Convergence issues in ABJ theories and validity of formula \eqref{eq:macdonald-rABJ}}}
\label{sec:convergence}

The rABJ generating function is defined by the matrix integral
\begin{equation}
 Z^{\mathrm{rABJ}}_{N,M}(q,t) = \frac{1}{N!M!}\int_{\mathbb{R}^{N+M}_+} \Delta_{q,t}(x,y) \mathe^{-V(x,y)}\prod_{i=1}^N\mathd x_i\prod_{a=1}^M\mathd y_a
\end{equation}
As it stands the integral is not convergent over the domain of integration $\mathbb{R}^{N+M}_+$ because of vanishing terms $(x_i-\sqrt{t/q}\,y_a)$ and $(y_a-\sqrt{t/q}\,x_i)$ in the denominator of the function $\Delta_{q,t}(x,y)$. A prescription therefore needs to be given to specify how to avoid such divergencies and make the integral well-defined. Analytical issues in $q,t$-deformed matrix integrals are usually hard to study, for this reason we restrict ourselves to the unrefined case $t=q$. This simplified version of the theory is already complex enough that we can use it as a toy model to explain the role of analytics in the definitions of the theory. The matrix integral then simplifies to
\begin{equation}
\label{eq:ABJ-like}
 Z^{\mathrm{ABJ}}_{N,M} = \frac{(-1)^{\binom{N+M}{2}}}{N!M!}\oint_{\mathcal{C}} \frac{\prod_{i<j}(x_i-x_j)^2\prod_{a<b}(y_a-y_b)^2}{\prod_{i,a}(x_i-y_a)^2}
 \mathe^{-\sum_i V(x_i)+\sum_a V(y_a)}\prod_{i=1}^N\mathd x_i\prod_{a=1}^M\mathd y_a
\end{equation}
where $\mathcal{C}$ is a contour to be determined and
\begin{equation}
 V(z) = \frac{\log^2 z}{2\log q} - \FI \log z
\end{equation}
Clearly, the choice of contour $\mathcal{C}$ depends on the behavior of the potential $V(z)$ but also on the poles in the denominator of the Cauchy determinant. Finding a contour compatible with a given choice of potential is not too hard however dealing with the poles can be subtle. For this reason we will assume that $V(z)$ is a generic potential for which a prescription for a contour $\mathcal{C}$ exists and we study how to modify such contour in such a way to avoid the poles while keeping the integral convergent.

In order to avoid divergencies along the contour $\mathcal{C}$ there are a couple of possibilities. One possibility is to regularize the divergent integral by cutting out a small region of the integration domain around the singularity. This type of regularization introduces boundaries in the integration domain which in turn are known to give rise to boundary terms in the ($q$-)Virasoro constraints \cite{Cassia:2021dpd}. Since, the constraint equations are no longer homogeneous one cannot expect the solution found in Section~\ref{sec:rABJ} to apply to this case. For this reason this type if regularization is not well-suited to our discussion.

Another possibility is to deform the original contour to some middle dimensional locus in $\mathbb{C}^{N+M}$ which picks up some of the poles in the integrand according to some prescription.

\begin{example}\label{ex:exampleABJ}
Consider $U(N+1|1)$ ABJ-like partition function
\begin{equation}
 Z^{\mathrm{ABJ}}_{N+1,1} = \frac{(-1)^{\binom{N+2}{2}}}{(N+1)!}\oint_{\mathcal{C}} \frac{\prod_{i<j}(x_i-x_j)^2}{\prod_{i}(x_i-y)^2} \mathe^{-\sum_{i=1}^{N+1}V(x_i)+V(y)}\prod_{i=1}^{N+1}\mathd x_i \mathd y
\end{equation}
with $\mathcal{C}$ defined in the following way: the integrals are ordered in such a way that we first take the residue at $y=x_{N+1}$, then the residue at $x_{N+1}=x_1$ and then all the other integrations in the variables $x_1,\dots,x_N$ along some contour $\mathcal{C}'$ which is middle-dimensional in $\mathbb{C}^N$. The $y$ and $x_{N+1}$ integrations can be carried out independently from all the others by using the identity
\begin{equation}
\label{eq:identityABJ}
\begin{aligned}
 &\oint_{x_{N+1}=x_1}\frac{\mathd x_{N+1}}{2\pi\mathi} \oint_{y=x_{N+1}}\frac{\mathd y}{2\pi\mathi} \,
 \frac{\prod_{1\leq i<j\leq N+1}(x_i-x_j)^2}{\prod_{i=1}^{N+1}(x_i-y)^2} \mathe^{-\sum_{i=1}^{N+1}V(x_i)+V(y)}
 =\\
 & = \oint_{x_{N+1}=x_1}\frac{\mathd x_{N+1}}{2\pi\mathi}
 \left(V'(x_{N+1})-2\sum_{i=1}^N\frac{1}{(x_{N+1}-x_i)} \right)\prod_{1\leq i<j\leq N}(x_i-x_j)^2\mathe^{-\sum_{i=1}^{N}V(x_i)} \\
 & = -2\prod_{1\leq i<j\leq N}(x_i-x_j)^2 \mathe^{-\sum_{i=1}^N V(x_i)}
\end{aligned}
\end{equation}
hence
\begin{equation}
 Z^{\mathrm{ABJ}}_{N+1,1} = -2(2\pi\mathi)^2\frac{(-1)^{\binom{N+2}{2}}}{(N+1)!}\oint_{\mathcal{C}'} \prod_{1\leq i<j\leq N}(x_i-x_j)^2 \mathe^{-\sum_{i=1}^{N}V(x_i)}\prod_{i=1}^{N}\mathd x_i
 = 2\frac{(2\pi\mathi)^2}{N+1}Z^{\mathrm{ABJ}}_{N,0}
\end{equation}
which we recognize as the hermitian matrix model partition function in rank $N$. Here $\mathcal{C}'$ can be chosen as usual by analyzing the potential $V(x)$, e.g. for $V(x)=x^2/2$ we can just take $\mathcal{C}'=\mathbb{R}^N$. 
\end{example}
The argument in the previous example can be generalized to any $U(N|M)$ and it provides evidence that indeed there is an equivalence with the bosonic matrix model for $U(N-M)$ as argued in Section~\ref{sec:rABJ} via $q$-Virasoro constraints.
This correspondence was already noticed in \cite{Okuda:2006fb} and \cite{Cassia:2021dpd} where classical Virasoro constraints were used to provide a proof of the equivalence of the supersymmetric and bosonic Gaussian matrix models.

In \cite{Vafa:2014iua} it was argued that this correspondence holds true at the ``perturbative'' level in the large $N_{\mathrm{eff}}$ limit however it should break down at finite  $N_{\mathrm{eff}}$ due to ``non-perturbative'' corrections. More explicitly, it was argued that the correlation functions of certain operators that vanish identically in the $U(N-M)$ model, do not vanish in the $U(N|M)$ dual, but rather they are of order $O(\mathe^{-a N_{\mathrm{eff}}})$. In the following we will show that the argument of \cite{Vafa:2014iua} does not apply to the integrals of the type \eqref{eq:ABJ-like} when the contour $\mathcal{C}$ is appropriately defined.

For concreteness we consider the case of the duality between $U(N+1|1)$ and $U(N)$. The operators considered in \cite{Vafa:2014iua} are multiples of the function
\begin{equation}
 e_{N+1}(x_1,\dots,x_{N})
\end{equation}
which is identically zero in the $U(N)$ theory by the definition of the elementary symmetric functions ($e_{k}(x_1,\dots,x_N)=0$ for $k>N$). Clearly, its matrix model expectation value will also be zero. The duality then implies that
\begin{equation}
 \left\langle se_{N+1}(x_1,\dots,x_{N+1},y)\right\rangle_{U(N+1|1)}
 = (\mathrm{const.})\times \left\langle e_{N+1}(x_1,\dots,x_{N})\right\rangle_{U(N)} = 0
\end{equation}
where, similarly to \eqref{eq:superMcdP}, $se_{N+1}$ is the supersymmetric version of $e_{N+1}$,
\begin{equation}
 se_{N+1}(x_1,\dots,x_{N+1},y)
 \equiv e_{N+1}\left(\left\{p_n=p_n(x,y)\right\}\right)
 = \prod_{i=1}^{N+1} (x_i-y)
\end{equation}
Therefore we need to show that even though $e_{N+1}(x_1,\dots,x_{N+1},y)$ is generically not zero, its expectation value is indeed zero. By definition, we can compute the expectation value as
\begin{equation}
 \left\langle \prod_{i=1}^{N+1} (x_i-y) \right\rangle_{U(N+1|1)}
 = \frac{(-1)^{\binom{N+2}{2}}}{(N+1)!}\oint_{\mathcal{C}} \frac{\prod_{i<j}(x_i-x_j)^2}{\prod_{i}(x_i-y)} \mathe^{-\sum_{i=1}^{N+1}V(x_i)+V(y)}\prod_{i=1}^{N+1}\mathd x_i \mathd y
\end{equation}
where we notice that the insertion reduces the degree of all the poles. Then, because of our choice of contour $\mathcal{C}$ as in Example~\ref{ex:exampleABJ} the integral vanishes. Observe that any other contour prescription will pick up some of the poles in the denominator of the measure, so that the argument still applies.

Similarly, we argue that insertions of $se_{N+1}(x_1,\dots,x_{N+1},y)^2$ also must vanish. The expectation value is given by
\begin{multline}
 \left\langle \prod_{i=1}^{N+1} (x_i-y)^2 \right\rangle_{U(N+1|1)}
 = \frac{(-1)^{\binom{N+2}{2}}}{(N+1)!}\oint_{\mathcal{C}'\cup\{x_{N+1}=x_1\}}
 \prod_{i=1}^{N+1} \mathd x_i \prod_{i<j}(x_i-x_j)^2 \mathe^{-\sum_{i=1}^{N+1}V(x_i)}
 \times \\ \times
 \oint_{y=x_{N+1}} \mathe^{V(y)} \mathd y
\end{multline}
so that the integral factorizes into two hermitian integrals of rank $N+1$ and $1$, respectively, and naively one would say that the expectation value is non-zero, however, upon closer inspection one realizes that the residues in $y$ and in $x_{N+1}$ both give zero because the poles that they were previously picking up have been canceled by the insertion of the operator $se_{N+1}^2$. This concludes our proof and shows that the correspondence between $U(N|M)$ and $U(N-M)$ matrix models holds exactly and not just ``perturbatively'' once a prescription for the choice of contour is given.

A thorough analysis of the refined case is beyond the scope of this paper however we expect that the general ideas outlined in this section will still hold. It then follows that the formula in \eqref{eq:macdonald-rABJ} is to be regarded as exact and does not receive non-perturbative corrections provided that the integral is appropriately regularized.

\providecommand{\href}[2]{#2}\begingroup\raggedright\endgroup

\end{document}